\documentclass[sigconf]{acmart}

\AtBeginDocument{%
  \providecommand\BibTeX{{%
    \normalfont B\kern-0.5em{\scshape i\kern-0.25em b}\kern-0.8em\TeX}}}

\copyrightyear{2022}
\acmYear{2022}
\setcopyright{acmcopyright}\acmConference[MSR '22]{19th International Conference on Mining Software Repositories}{May 23--24, 2022}{Pittsburgh, PA, USA}
\acmBooktitle{19th International Conference on Mining Software Repositories (MSR '22), May 23--24, 2022, Pittsburgh, PA, USA}
\acmPrice{15.00}
\acmDOI{10.1145/3524842.3528469}
\acmISBN{978-1-4503-9303-4/22/05}
    
\usepackage[utf8]{inputenc}
\usepackage{url}
\usepackage{hyperref}
\usepackage{todonotes}
\usepackage{listings}
\usepackage{tabularx}
\usepackage{caption}
\usepackage{xcolor,colortbl} 
\usepackage{flushend}
\title{WeakSATD: Detecting Weak Self-admitted Technical Debt}
\usepackage{natbib}
\usepackage{graphicx}
\usepackage{enumitem}
\usepackage{tcolorbox}
\DeclareRobustCommand{\summarybox}[1]{%
\begin{tcolorbox}[
    boxrule=.5pt,
    left=3pt,
    right=3pt,
    top=3pt,
    bottom=3pt,
    colback=black!5!white,
    colframe=black!75!black
    ]
    #1
\end{tcolorbox}
}

\author{Barbara Russo}
\orcid{0000-0003-3737-9264}
\affiliation{%
  \institution{Free University of Bozen-Bolzano}
  \country{Italy}
}
\email{barbara.russo@unibz.it}

\author{Matteo Camilli}
\orcid{0000-0003-2491-5267}
\affiliation{%
  \institution{Free University of Bozen-Bolzano}
  \country{Italy}
}
\email{matteo.camilli@unibz.it}

\author{Moritz Mock}
\affiliation{%
  \institution{Free University of Bozen-Bolzano}
  \country{Italy}
}
\email{moritz.mock@stud-inf.unibz.it}

\definecolor{LightGray}{gray}{0.9}

\begin{document}

\lstset{breaklines=true,
basicstyle=\footnotesize\ttfamily,
keywordstyle=\color{blue},
language=C
}

\captionsetup[lstlisting]{labelfont={sc} }
\begin{abstract}
Speeding up development may produce technical debt, i.e., not-quite-right code for which the effort to make it right increases with time as a sort of interest. 
Developers may be aware of the debt as they  admit it in their code comments.   Literature reports that such a self-admitted technical debt survives for a long time in a program, but it is not yet clear its impact on the quality of the code in the long term. 
We argue that self-admitted technical debt contains a number of different weaknesses that may affect the security of a program. Therefore, the longer a debt is not paid back the higher is the risk that the weaknesses can be exploited. 
To discuss our claim and rise the developers' awareness of the vulnerability of the self-admitted technical debt that is not paid back, we explore the self-admitted technical debt in the Chromium C-code to detect any known weaknesses. In this preliminary study, we first mine the Common Weakness Enumeration repository to define heuristics for the automatic detection and fix of weak code. Then, we parse the C-code to find self-admitted technical debt and the code block it refers to.  Finally, we use the heuristics to find weak code snippets associated to self-admitted technical debt and recommend their potential mitigation to developers. Such knowledge can be used to prioritize self-admitted technical debt for repair. 
A prototype has been developed and applied to the Chromium code. Initial findings report that 55\% of self-admitted technical debt code contains weak code of 14 different types.
\end{abstract}

\begin{CCSXML}
<ccs2012>
   <concept>
       <concept_id>10011007.10011074.10011111.10011696</concept_id>
       <concept_desc>Software and its engineering~Maintaining software</concept_desc>
       <concept_significance>300</concept_significance>
       </concept>
       <concept>
    <concept_id>10002978.10003022</concept_id>
    <concept_desc>Security and privacy~Software and application security</concept_desc>
    <concept_significance>300</concept_significance>
    </concept>
 </ccs2012>
\end{CCSXML}

\ccsdesc[300]{Software and its engineering~Maintaining software}
\ccsdesc[300]{Security and privacy~Software and application security}

\keywords{Self-admitted technical debt, Weak code, Security, Vulnerability}

\maketitle

\section{Introduction} \label{sec:intro}

Not-quite-right code is introduced for short-term needs~\cite{Fowler1999,Cunningham2009}.  If it is not fixed, it may increase over time with negative impact on code quality~\cite{ZazworkaEtAl2011} and, if it is fixed, it may cause cost of additional rework~\cite{PotdarShihab2014}. Such additional effort is called \emph{technical debt}.
Examples of technical debt are code smells and bug hazards.
Identifying automatically such a code is the goal of recent lines of research~\cite{ZazworkaEtAl2011,ZampettiEtAl2018,BavotaRusso2016}. One option is to trace \textit{Self-Admitted Technical Debt} (SATD), i.e., code comments explicitly introduced by developers to tag the technical debt~\cite{LiuEtAl2018,BavotaRusso2016, PotdarShihab2014}. 
Such comments can be automatically detected and the associated code can be identified and removed or modified to mitigate the debt (i.e., paying back technical debt)~\cite{ZampettiEtAl2018}.
According to developers' opinion, SATD is not introduced because of pressure~\cite{BavotaRusso2016}, but it is rather included intentionally to track future bugs and areas of
the code that need refactoring~\cite{MaldonadoEtAl2017a, LimEtAl2013}. It has been hypothesised that SATD may affect  software correctness~\cite{LimEtAl2013}, but literature has not yet reported a conclusive answer~\cite{SierraEtAl2019, ZampettiEtAl2018,BavotaRusso2016, WehaibiEtAl2016}. 
Not-quite-right code may indeed contain \emph{weaknesses} (i.e., code that exposes software to security breaches). Such weaknesses can be exploited by a party to cause the product to modify or access unintended data, interrupt proper execution, or perform incorrect actions that were not specifically granted to the party who exploits the weakness. Thus, the longer this code remains in the software, the higher will be the interest to pay it back, and the longer will be the exposure of the software to third party's exploitation. As SATD on average survives for  over 1,000 commits~\cite{BavotaRusso2016}, weaknesses in the  SATD-related code  represent a real security risk. Even though developers are aware of the technical debt since they self-admit it, they may not be aware of the portion of such debt that is also weak and the security risks at which their software is exposed.
In this work, we consider the following research question:

\begin{enumerate}[start=1,label={\bfseries RQ:},leftmargin=1.1cm]
\item Is self-admitted technical debt related to weaknesses in source code?
\end{enumerate}

We argue that \emph{detecting weaknesses associated to SATD can increase developers' awareness on the vulnerability of their code and on the risk of not paying the technical debt back and help them  plan  maintenance activities for security concerns (e.g., prioritize SATD for repair).}

In this paper, we present a preliminary study in which we analyzed the source code written in C of the Chromium project~\cite{chromium} to understand whether code blocks to which SATD comments refer may contain weaknesses.
To achieve this goal, we developed the \emph{WeakSATD} approach that  mines the public Common Weaknesses and Enumeration (CWE) repository~\cite{cwe} to derive a set of heuristics to detect known weaknesses in software code and recommend their mitigation. 
SATD comments and related code are then retrieved. Finally, the heuristics are  used to automatically detect the presence and the types of weaknesses in the code to which SATD comments refer. Mitigation to such weaknesses are then recommended.
We believe that such knowledge can be leveraged to prioritize SATD for repair (e.g., with or without weaknesses or with one or more weaknesses)   and speed up SATD removal.

The remainder of this paper is as follows.
We introduce \emph{WeakSATD} in Sec.~\ref{sec:weaksatd}.
We discuss our initial findings with the Chromium project in Sec.~\ref{sec:eval}.
We briefly report related work in Sec.~\ref{sec:related}.
We conclude the paper and we present future directions in Sec.~\ref{sec:conclusion}.

\begin{figure*}
    \centering
    \includegraphics[width=.65\textwidth]{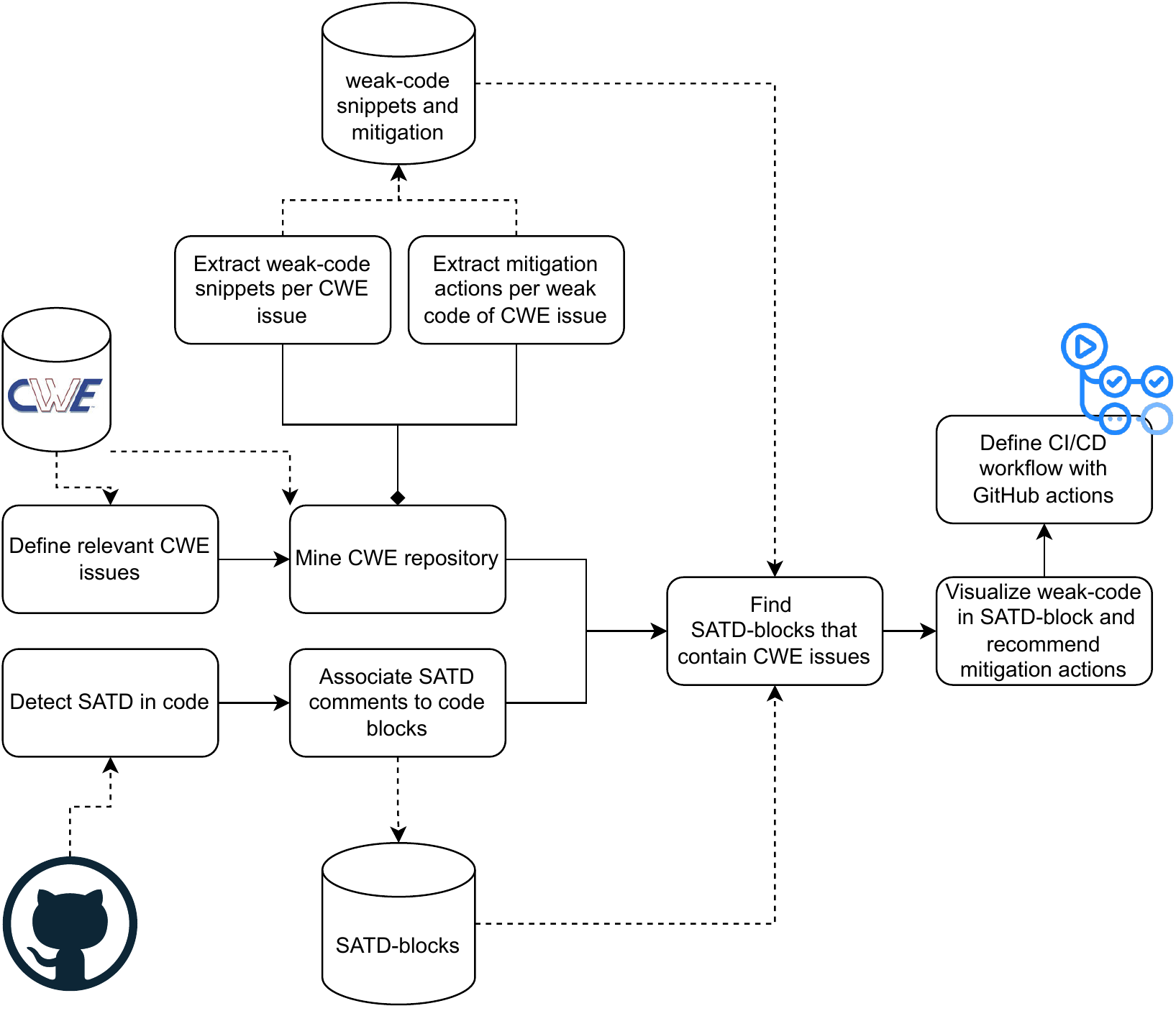}
    \caption{Overview of our approach to detect weaknesses in code associated to SATD.}
    \label{fig:overview}
\end{figure*}

\section{WeakSATD}\label{sec:weaksatd}

In this section, we introduce an overview of the whole approach (Sec.~\ref{sec:overview}), we describe how we mine and leverage the information in the CWE issues (Sec.~\ref{sec:miningCWE}),  how we derive our weakness heuristics (Sec.~\ref{sec:heuristics}) used to analyze the Chromium project, the detection of SATD-blocks (Sec.~\ref{sec:SATD_Methods}), and the WealSATD prototype (Sec.~\ref{sec:weak-satd}).

\subsection{Overview}\label{sec:overview}

Figure~\ref{fig:overview} illustrates a high-level overview of our approach that includes the following steps.

\emph{Define relevant CWE issues}. We first mine the CWE repository that contains the state-of-the-art  list of existing weaknesses (hardware and software) and extract the weakness types (called \emph{CWE issues}) that pertain to software code written in C and include in their description of some C-code examples. 

\emph{Mine CWE repository}. Then, we define weakness heuristics as regex rules in C by analyzing the code snippets reported as \emph{weak-code examples} in these CWE issues' description (we call them \emph{weak-code snippets}). For the rest of the paper, we refer to \emph{weak code} as a portion of the code that contains one or more weak-code snippets. 
\emph{Detect SATD in code}. At the same time, we determine the \emph{SATD-blocks} as the blocks (e.g., method block or block of a loop) in the C-code of Chromium to which SATD instances refer. To this aim, we parse the code to identify SATD comments with regex rules matching the 62 patterns defined by  Potdar and Shihab~\cite{PotdarShihab2014} and to detect code blocks associated with them.

\emph{Find SATD-blocks that contain CWE issues}. Finally, we use  regex rules to find weak-code snippets in the SATD-blocks and provide mitigation actions by exploiting the information contained in the description of the CWE issues. The same information is leveraged to provide mitigation actions. 

We have developed a prototype tool that automatizes our approach, visualizes instances of SATD-blocks that contain potential weaknesses, and suggests mitigation actions\cite{zenodo}. 
Finally, we implemented the approach also using GitHub Actions~\cite{githubActions} to enable automation of our approach in CI/CD pipelines.
A more detailed description of our approach follows in the next sections.

An anonymized package containing the implementation of \emph{WeakSATD} demo, the list of heuristics, the Github Actions, and the data extracted from CWE is publicly available~\cite{zenodo}.

\subsection{Exploring the CWE repository}\label{sec:miningCWE}
CWE is part of a larger MITRE~\cite{mitre} initiative for collecting, classifying, and publishing data on  weaknesses, vulnerabilities, and attacks to software and hardware.  
CWE is a public community-maintained moderated repository of over 900 types of software and hardware known weaknesses. 
Weakness types are entered in CWE as issues (\emph{CWE issue}s). Each issue may include a description, relationships with other issues (if any), platforms and programming languages that can be affected, known consequences related to attacks or system malfunctioning, demonstrative code examples, and potential mitigation actions as in Table~\ref{tab:CWEInfo}.  
\begin{table}[b]
    \centering
    \scriptsize
    \caption{Relevant information in a CWE issue.}
    \label{tab:CWEInfo}
    \begin{tabular}{@{}lp{0.6\columnwidth}}
    \toprule
    \textit{Title}& \textbf{CWE-242: Use of Inherently Dangerous Function}\\
    \midrule
    \textit{Description}& The program calls a function that can never be guaranteed to work safely \\
   \textit{Extended Description}& Certain functions behave in dangerous ways regardless of how they are used. ... The gets() function is unsafe because does not perform bounds checking on the size of its input. An attacker can easily send arbitrarily-sized input to gets() and overflow the destination buffer. ...\\
   \textit{Relationships}& Relevant to the view "Software Development" (CWE-699)\\
  &\begin{tabular}{p{30pt}p{10pt}p{10pt}p{50pt}}\hline
    \textit{Nature}& 	\textit{Type}  & \textit{ID}&  \textit{Name} \\
    MemberOf& C&1228&API/Function Errors\\
    \hline
   \end{tabular}\\
   \textit{Modes of Introduction}:& Phase: Implementation\\
   \textit{Applicable Platform}&Languages: C, C++\\
   \textit{Common Consequences}&Technical Impact: Varies by \textit{Context}\\
   \textit{Likelihood Of Exploit}&High\\
   \textit{Demonstrative Examples}&
The code below calls gets() to read information into a buffer.\\
&Example. Language: C 
\begin{lstlisting}[belowskip=-0.8 \baselineskip,basicstyle=\ttfamily\scriptsize]
  char buf[BUFSIZE];
  gets(buf);
\end{lstlisting}
\\
\textit{Potential mitigations}&Phases: Implementation; Requirements
ban the use of dangerous functions. Use their safe equivalent.\\
&Phase: Testing;
Use grep or static analysis tools to spot usage of dangerous functions.\\
\bottomrule
\end{tabular}
\end{table}
Search in the CWE repo can be performed by  issue-ID or keywords. No API is available, and the dataset can be downloaded as a set of $\mathtt{.cvs}$ or $\mathtt{.hml}$ files. 
CWE issues are also linked to records stored in the Common Vulnerabilities and Enumerations (CVE)~\cite{cve} or National Vulnerabilities Database (NVD)~\cite{nvd}, i.e., the U.S. government repository of standards-based vulnerability management data represented using the Security Content Automation Protocol (SCAP). CVE contains data on vendors, products and versions of products as well as vulnerabilities per type (e.g., Denial of Service, Code Execution, Overflow, etc.). Finally, the exploitation of vulnerabilities (e.g., exploitation code and information) are maintained in the exploit database~\cite{exploitDB}.
Entries in the exploit DB,  CVE/NVD and CWE repositories are all linked through the identifier \emph{CWE-ID}. Some bug tracking systems  (e.g., Mozilla~\cite{bugzilla}) are also using the CWE-ID to annotate issues related to security. 
Thus, by linking SATD to CWE issues, \emph{we can trace technical debt to weaknesses, their effects (vulnerabilities and bugs) and possible attacks (exploitation).}

The CWE repository contains more than 920 different weaknesses types (CWE issues) for software and hardware linked each other through their CWE-ID when this is specified in the  \emph{Relationships} field, as shown in Table~\ref{tab:CWEInfo}. 
For instance, CWE-242 is a ``member of'' the general category CWE-699. The relationships field can include different types of associations between CWE issues like MemberOf, ChildOf and ParentOf.

To  select the relevant list of CWE issues, we have inspected the 900 CWE types also exploiting the relationship they declare in their description. We have then manually isolated the issues that: (1) pertain only to software development (total 419), (2) can be found in the C programming language, and (3) contain a code example in the C language.  
By applying these criteria, we finally obtained a sample of 80 CWE issues.

\subsection{Weakness Heuristics} \label{sec:heuristics}
To identify weaknesses in the code, we define a set of  heuristics  by manually  inspecting the 80 CWE issues. Firstly, we further reduce the number of CWE issues to analyze by exploiting the dependency structures among them. Specifically, we re-use the heuristic of the parent node when possible and we implement only the heuristics for the child nodes when the parent's issue description is too coarse to detect its children.  
At the end, we come up with \emph{34 relevant heuristics}.
After applying the aforementioned selection criteria, we implement regex rules to parse the  C-code by  analyzing the description and the  code snippets included as examples of vulnerability in C in the description of the corresponding CWE issue.   The recommended mitigation actions of the CWE issue are further extracted from the CWE description and associated to the weak-code snippet. 
The heuristics we find depend on the specific programming language, the available information in the CWE repository, and researchers' personal knowledge of the language, although they are independent from the Chromium project. Thus, they are far from being a complete set for any C-written project  and represent only a demonstrative sample to illustrate our research idea.
Table~\ref{tab:CWE676} illustrates an example of heuristic we defined for the issue CWE-676 in the C language.  The weak-code snippet in CWE-676  creates a local copy of a buffer to perform some data manipulations.
The function $\texttt{strcpy()}$ copies a  string into a buffer with no control on the size of the string potentially exposing the software to buffer overflow exploits. 
The heuristic we defined searches for  the  potentially dangerous functions in C, $\texttt{strcpy()}$ illustrated in the CWE-676 example and also for each of  the potentially vulnerable functions in the banned list maintained by Microsoft~\cite{Howard}. 
\begin{table}
\caption{A heuristic for CWE-676.}
    \label{tab:CWE676}
        \scriptsize
\begin{tabular}{@{}lp{0.8\columnwidth}}
    \toprule
    \textit{Title}& \textbf{CWE-676: Use of Potentially Dangerous Function}\\
    \midrule
     \textit{Example in C}&\begin{lstlisting}[basicstyle=\ttfamily\scriptsize, belowskip=-0.8 \baselineskip, aboveskip=-0.5 \baselineskip] 
    void manipulate_string(char * string){
       char buf[24];
       strcpy(buf, string);
       ...
    }
    \end{lstlisting} \\
    \textit{Heuristic}& Search for ``$\mathtt{strcpy(}$'' with no alphanumerical character in front of the keyword \\ 
  \bottomrule
\end{tabular}
\end{table}
More complex regex rules have been defined in other cases. 
The set of regex rules we implemented are available in the replication package~\cite{zenodo}.

\subsection{SATD-blocks}
\label{sec:SATD_Methods}
As per common practice~\cite{BavotaRusso2016,ZampettiEtAl2018},  we implement a set of regex rules to search among all comments for the  62 patterns defined by  Potdar and Shihab~\cite{PotdarShihab2014} to detect SATD in code~\cite{SATDpatterns} and use \textsc{srcML} to decompose C-code in comments and blocks,~\cite{BavotaRusso2016,ZampettiEtAl2018, PotdarShihab2014}. 
Then, we associate a SATD instance to a block following a proximity rule as we derived as in the following.  
Firstly, in all comments in the code we search the ones that include one or more  SATD  patterns (\emph{SATD-comments}). Secondly,  we consider all code blocks and associate each of them to a SATD comment by a proximity rule based on the results of a  Wilcoxon test between the two distributions of lines of code between SATD comments and enclosing or following code blocks. For instance, for the Chromium project, we decided to select blocks that  occur \emph{after} a comment since the Wilcoxon test significantly ($\alpha$=0.05) reported the block after the comment as the nearest one\footnote{Note this result may be different in other software projects.}.
SATD comments that are not associated to any code block are not included in the analysis. 
We finally call \emph{SATD-block}, the code block associated in this way to a SATD comment.

\subsection{Weak SATD} \label{sec:weak-satd}
With the regex rules defined from the heuristics, we search weak-code snippets in SATD-blocks and called \emph{Weak SATD} the positive results. 
It is worth noting that a block can contain more instances of such snippets from the same or different CWE issues, Listing~\ref{cd:SATDVulnerable}.
For each of the found weak-code snippet, we  also retrieve the mitigation actions recommended by the CWE issue in the  field ``potential mitigations'', as shown in Table~\ref{tab:CWEInfo}. %

\noindent\paragraph{Prototype}\label{pr:zenodo}
We developed a  prototype tool implementing the \emph{WeakSATD} approach. The tool is  a web-based application that uses react.js and  npm-packages (e.g., \texttt{lodash}  and  \texttt{adm-zip}). 
A RESTful API is implemented by using \texttt{express} and \texttt{mongoose} to interact with \texttt{MongoDB} and \texttt{babel} to allow compliance with the  react  syntax. The API handles  saving of relevant CWE data and statistics on weak-code snippets in MongoDB as JSON record.  The tool uses \texttt{SrcML} to retrieve relevant items from the code (comments, blocks). 
Finally, we developed GitHub Actions for the analysed CWE issues to exemplify the integration of \emph{WeakSATD} into CI/CD pipelines. 
The action is executed upon new release commits to spot weak-code snippets in SATD-blocks. Then it notifies the developer with potential mitigation actions.
The GitHub action runs as a job configured through a \texttt{.yml} file as illustrated in Listings~\ref{lst:action}.
\begin{lstlisting}[label="lst:action", caption=A GitHub Action for \emph{WeakSATD}., label=lst:action, basicstyle=\ttfamily\small]
jobs:
  weakSATDAction:
    runs-on: ubuntu-latest
    steps:
      - uses: actions/checkout@v2
      - name: Run weakSATD Action
        uses: <account>/WeakSATD-Action@v0.0.1
\end{lstlisting}

\section{Initial findings in Chromium} \label{sec:eval}
We applied our approach to the C sources of Chromium~\cite{chromiumRepo}, version $88.0.4323$. We selected this project as it was originally used by Nord \emph{et al.}~\cite{NordEtAl2016} for their initial findings on the relation between vulnerabilities and technical debt and by Potdar and Shihab~\cite{PotdarShihab2014} in the definition of the 62 SATD patterns.

By mining the CWE repository, with our selection criteria (Sec.~\ref{sec:weaksatd}), we obtained and implemented $34$ weakness heuristics.
For each CWE issue, we refined the initial implementation of the weakness heuristics in our demo tool after manually analyzing a sample of $775$ files sampled with $90\%$ confidence level and $5\%$ margin of error. %(population proportion sampling size).  
With this manual inspection, we also found that the SATD pattern ``take care''  produced quite a few false positives in the SATD comments of Chromium as illustrated in Listing \ref{lst:fptakecare}. To stay on the conservative side, we did not consider this pattern in our analysis. 

\begin{lstlisting}[language=C, caption={False positive comment for the ``take care'' SATD pattern.},label={lst:fptakecare},basicstyle=\ttfamily\small]
// Callers are encouraged to use the setters provided which take care of setting |options| as desired.
\end{lstlisting}

Finally, we made sure that for this sample, the tool predicted weak-SATD blocks that were actual weak-code according to the CWE description. 
We applied our weakness heuristics and our SATD-block rule to 41753 C files. Table~\ref{tab:statistics} shows that the percentage  of files with SATD and the percentage of comments with SATD found with our approach is within the ranges reported in literature.  
\begin{table}[h]
\centering \scriptsize
\caption{\% SATD files and comments in recent literature.}
\label{tab:statistics}
\begin{tabular}{lcc}
\toprule
&\% files with SATD comments & \% SATD comments \\
\midrule
\rowcolor{LightGray}WeakSATD~\cite{zenodo}&1.6\%$^*$&0.1\%$^*$\\
Potdar et al.~\cite{PotdarShihab2014}  &2.4-31\%&-\\
Bavota et al.~\cite{BavotaRusso2016} &-&0.2-0.4\% \\
Maldonado et al.~\cite{MaldonadoEtAl2017a}&-&0.02-0.21\%\\ 
Zampetti et al.~\cite{ZampettiEtAl2018}&-&0.02-0.21\%\\ 
Iammarino et al.~\cite{Iammarino2019ICSME}&-&0.02-0.21\%\\ 
Fucci et al.~\cite{Fucci2021MSR}&  -&0.02-0.21\%\\           
\bottomrule
\end{tabular}
\flushleft\footnotesize{$^*$Computed considering SATD comments with blocks only.}
\end{table}

% \begin{table}[hbt]
% \centering \scriptsize
% \caption{\% SATD  and \% CWE weak-code.}
% \label{tab:statistics}
% \begin{tabular}{lll|ll}
% \toprule
% & \multicolumn{2}{c}{files} & \multicolumn{2}{c}{comments} \\ \midrule  & with weak-code & with SATD & SATD & weak SATD \\ \midrule
% \rowcolor{LightGray}\emph{WeakSATD}~\cite{zenodo}  & 26\%   & 1.6\%$^*$    & 0.1\%$^*$ & 55\%  \\
% Potdar et al.~\cite{PotdarShihab2014}                                                                                                                                                                                                          & -              & 2.4-31\% & -                               & -         \\
% \rowcolor{LightGray}Bavota et al.~\cite{BavotaRusso2016}                                                                                                                                                                                                           & -              & -                            & 0.2-0.4\%   & -         \\
% \begin{tabular}[c]{@{}l@{}}Maldonado et al.~\cite{MaldonadoEtAl2017a}\\ Zampetti et al.~\cite{ZampettiEtAl2018}\\ Iammarino et al.~\cite{Iammarino2019ICSME}\\ Fucci et al.~\cite{Fucci2021MSR}\end{tabular} & -              & -                            & 0.02-0.21\% & -         \\ \bottomrule
% \end{tabular}
% \flushleft\footnotesize{$^*$Computed considering SATD comments with blocks only.}
% \end{table}
% % %
We found that 10885 files ($26\%$ of all C files) contain a number of  weak-code snippets  ranging from 0 for issue CWE-243 to 4091 for issue CWE-783~(average=1197, Q1=14, median Q2=88, Q3=365). It is worth noting that issue CWE-783 is very common as it refers to the use of an expression in which operator precedence causes incorrect logic.  
We also found 847 distinct SATD blocks for about $0.1\%$ of all comments. 
We found  SATD blocks 
in 634 different files (about $1.6\%$ of all files).
Out of those SATD blocks, 465, that is, $55\%$ of them contain at least one potential weakness according to the CWE catalog. 
We found that weak SATD-blocks are  distributed over 14 different CWE issues (about $41\%$ of the implemented issues) with a maximum of 197 SATD-blocks for issue CWE-483 ``Incorrect Block Delimitation" (e.g., missing brackets for an ``if statement'').

Our initial findings with the Chromium project have no ambition of generalization also because, at the current stage, we do not cover all CWE issues.
However, we can summarize our findings as follows to answer our initial research question.

\summarybox{\textbf{RQ summary}:
In our experiments with the Chromium project, we found that $26\%$ of all source files contain at least one weak-code snippet.
Furthermore, $55\%$ of the SATD blocks contain weak-code snippets of 14 different CWE issues.
%According to our experience, the likelihood of finding weaknesses in SATD is high.
Indeed, correlation between weaknesses and SATD warrant further research.
}

\begin{lstlisting}[caption={SATD comment and relative block with multiple CWE instances.}, xleftmargin=5.0ex, numbers=left,basicstyle=\ttfamily\small,label=cd:SATDVulnerable,linewidth=\columnwidth]
/* FIXME: this code assumes that sigmask is an even multiple of the size of a long integer. */ 

unsigned long *src = (unsigned long const *) set;
unsigned long *dest = (unsigned long *) &(thread.p->sigmask);

switch (how)
{
    case SIG_BLOCK:
    for (i = 0; i < (sizeof (sigset_t) / sizeof (unsigned long)); i++)
    {
      /* OR the bit field longword-wise. */
      *dest++ |= *src++;
    }
    break;
    case SIG_UNBLOCK:
    for (i = 0; i < (sizeof (sigset_t) / sizeof (unsigned long)); i++)
    {
      /* XOR the bitfield longword-wise. */
      *dest++ ^= *src++;
    }
    case SIG_SETMASK:
    /* Replace the whole sigmask. */
    memcpy (&(thread.p->sigmask), set, sizeof (sigset_t));
    break;
}
\end{lstlisting}

The Chromium code in Listing \ref{cd:SATDVulnerable} shows an example of SATD comment and its SATD block in which our tool detects the following instances of CWE-issues:
CWE-478 ``Missing Default Case in Switch-statement'' (at the end of the switch), CWE-484 ``Omitted Break Statement in Switch'' (after the second case), CWE-242/676 ``Use of Potentially/Inherently Dangerous Function'' (line 23).

\section{Related work} \label{sec:related}

In our work we adopt a simple pattern-based SATD detection method, while other existing approaches may be in principle adopted to achieve higher accuracy~\cite{Ren2019TOSEM,ZYu2020TSE}.
Supervised approaches may be more precise in detecting SATD instances ($60-85\%$ precision with ML~\cite{ZYu2020TSE} vs. $75\%$ with unsupervised search~\cite{BavotaRusso2016}).
Nevertheless, they require ground truth and training effort that is not needed in pattern-based SATD detection.

% Recent methods to identify vulnerabilities in software projects include  Deep Learning or Machine Learning approaches (with text mining features), to predict vulnerabilities~\cite{StuckmanEtAl2017,LiFSE2021,mazuerarozoEtAl2021}.
% These models are usually built by mining the history of software code ~\cite{SabettaBezzi2018}, by mining product-independent vulnerability repositories like CWE~\cite{GkortzisEtAl2018}, or by combining the two approaches~\cite{NordEtAl2016,GkortzistAl2018,NeuhausEtAl2007}.

Considering the detection (or prediction) of security threats, recent approaches exploit the information in the  history of software projects and connecting it to software vulnerabilities, \cite{LiFSE2021,mazuerarozoEtAl2021,SabettaBezzi2018,GkortzisEtAl2018,StuckmanEtAl2017,NordEtAl2016,NeuhausEtAl2007}. For instance, Mazuera-Rozo \emph{et al.}~\cite{mazuerarozoEtAl2021} classify code functions with Deep and Machine Learning techniques  to classify them by known vulnerabilities, such as deadlock, race condition, and null pointer dereference.
The information on software vulnerabilities is used to label the categories of the classifiers and the focus is on vulnerabilities (e.g., buffer overflow) and not on software weaknesses (e.g., inadequate container capacity).

Table~\ref{tab:literature} lists recent pieces of work explicitly linking their results to a set of CWE issues.
As shown in the table, there is little overlapping with the issues considered in \emph{WeakSATD}.
Our approach is indeed different. %
It exploits the information in the weakness repository (CWE) to detect weak code in software projects.  

The relationship between technical debts and weaknesses has been recently studied~\cite{NordEtAl2016,siavvas2019empirical}.
Initial findings confirm software developers use technical debt concepts to discuss design
limitations and their consequences.
However, correlations between vulnerabilities and technical debt indicators requires further research.
The approach in~\cite{siavvas2019empirical} uses security bugs in issue tracking systems to identify vulnerabilities. Such bugs are not weaknesses but errors in code. According to the MITRE definition, weaknesses are code snippets that can be potentially (not yet) be exploited. Different from bugs and vulnerabilities, weaknesses represent, in our vision, a sort of debt.  None of the aforementioned approaches aim at detecting SATD or connect SATD (or TD) with weaknesses or exploit the information in CWE to identify weaknesses as \emph{WeakSATD}.

\begin{table}[htb] \caption{Recent literature, number of CWE issues analysed and CWE-isse in common with \emph{WeakSATD}.}
    \label{tab:literature}
    \centering \scriptsize
    \begin{tabularx}{240pt}{lcX}
    \toprule
    Datasets & CWE issues &CWE issues in common with \emph{WeakSATD}\\
    \midrule
        \rowcolor{LightGray}Shallow-Deep (2021)  \cite{mazuerarozoEtAl2021}& N/A & N/A \\
        Juliet Test Suite (2018) \cite{russellEtAl2018,julietEtAl} & 118 & 195, 196, 401, 415, 416 \\
        \rowcolor{LightGray}DRAPER VDISC (2018) \cite{russellEtAl2018,VDISC} &4&none\\
       Zou et al. (2019)\cite{ZouEtAl2019} &33& 467, 676\\ 
       \rowcolor{LightGray}WeakSATD ~\cite{zenodo}& 34  &-\\
       \bottomrule
    \end{tabularx}
\end{table}

\section{Conclusion and future Directions} \label{sec:conclusion}
More research is needed to understand the relation between vulnerabilities and technical debt~\cite{NordEtAl2016}. 
We believe that \emph{WeakSATD} represents a novel contribution  in this direction.   
We applied \emph{WeakSATD} to the Chromium project that has been previously used as test bed in research on technical debt (e.g., \cite{PotdarShihab2014, NordEtAl2016}). We found that more than 55\% of  SATD instances contain weak code of 34 different CWE issues. 
It is worth noting that the weak SATD report must be taken as a warning only as some of the found weak code snippets may be perfectly fine in some circumstances.
At the current stage, our work is not meant to predict vulnerabilities or attacks with high precision or recall, but it can be used by developers to prioritize and boost the repair of SATD with the help of the recommended mitigation. 
In addition, our initial findings with the Chromium project have no ambition of generalization especially because we were able to define a substantial amount if weakness heuristics but still not covering all CWE issues. This allowed us to better focus on their definition and implementation, but it prevented us from being more extensive with our findings. 
Our plan for future work, is to explore other publicly available datasets that are connected with the CWE issues (e.g., the Juliet Test Suite~\cite{julietEtAl}) and can be used to define additional heuristics. 
We also plan to exploit the whole chain of information available in MITRE (or MITRE-linked repositories) and provide an instrument that recommend developers on the whole chain of consequences (e.g., vulnerabilities, exposure, bugs, and attacks) of leaving SATD in code. 
The CWE indexing can be used for this scope, but not only. Indeed, some projects do not encode the CWE-ID in their bug issues and security issues can be traced in other ways, like regexes in commit messages, such as the approach introduced in~\cite{NordEtAl2016}. 
%To increase the perceived usefulness of \emph{weakSATD}, it is also important avoiding false alarms. 
We also plan to study the perceived usefulness of \emph{weakSATD} by involving professional developers in controlled experiments with humans.
%interview developers and investigate to what extent \emph{weakSATD} can be considered  useful for developers. 

\bibliographystyle{ACM-Reference-Format}
\bibliography{references}

%%% -*-BibTeX-*-
%%% Do NOT edit. File created by BibTeX with style
%%% ACM-Reference-Format-Journals [18-Jan-2012].

\begin{thebibliography}{39}

%%% ====================================================================
%%% NOTE TO THE USER: you can override these defaults by providing
%%% customized versions of any of these macros before the \bibliography
%%% command.  Each of them MUST provide its own final punctuation,
%%% except for \shownote{}, \showDOI{}, and \showURL{}.  The latter two
%%% do not use final punctuation, in order to avoid confusing it with
%%% the Web address.
%%%
%%% To suppress output of a particular field, define its macro to expand
%%% to an empty string, or better, \unskip, like this:
%%%
%%% \newcommand{\showDOI}[1]{\unskip}   % LaTeX syntax
%%%
%%% \def \showDOI #1{\unskip}           % plain TeX syntax
%%%
%%% ====================================================================

\ifx \showCODEN    \undefined \def \showCODEN     #1{\unskip}     \fi
\ifx \showDOI      \undefined \def \showDOI       #1{#1}\fi
\ifx \showISBNx    \undefined \def \showISBNx     #1{\unskip}     \fi
\ifx \showISBNxiii \undefined \def \showISBNxiii  #1{\unskip}     \fi
\ifx \showISSN     \undefined \def \showISSN      #1{\unskip}     \fi
\ifx \showLCCN     \undefined \def \showLCCN      #1{\unskip}     \fi
\ifx \shownote     \undefined \def \shownote      #1{#1}          \fi
\ifx \showarticletitle \undefined \def \showarticletitle #1{#1}   \fi
\ifx \showURL      \undefined \def \showURL       {\relax}        \fi
% The following commands are used for tagged output and should be
% invisible to TeX
\providecommand\bibfield[2]{#2}
\providecommand\bibinfo[2]{#2}
\providecommand\natexlab[1]{#1}
\providecommand\showeprint[2][]{arXiv:#2}

\bibitem[Bavota and Russo(2016)]%
        {BavotaRusso2016}
\bibfield{author}{\bibinfo{person}{Gabriele Bavota} {and}
  \bibinfo{person}{Barbara Russo}.} \bibinfo{year}{2016}\natexlab{}.
\newblock \showarticletitle{A Large-scale Empirical Study on Self-admitted
  Technical Debt}. In \bibinfo{booktitle}{\emph{Proceedings of the 13th
  International Conference on Mining Software Repositories}} (Austin, Texas)
  \emph{(\bibinfo{series}{MSR '16})}. \bibinfo{publisher}{ACM},
  \bibinfo{address}{New York, NY, USA}, \bibinfo{pages}{315--326}.
\newblock
\showISBNx{978-1-4503-4186-8}
\urldef\tempurl%
\url{https://doi.org/10.1145/2901739.2901742}
\showDOI{\tempurl}


\bibitem[Chromium(2022)]%
        {chromium}
\bibfield{author}{\bibinfo{person}{Chromium}.} \bibinfo{year}{2022}\natexlab{}.
\newblock \bibinfo{title}{Chromium project}.
\newblock \bibinfo{howpublished}{\url{https://www.chromium.org/Home}}.
\newblock
\newblock
\shownote{Last accessed Jan. 2022}.


\bibitem[Corporation(2022)]%
        {mitre}
\bibfield{author}{\bibinfo{person}{MITRE Corporation}.}
  \bibinfo{year}{2022}\natexlab{}.
\newblock \bibinfo{title}{Federally Funded Research and Development Centers}.
\newblock \bibinfo{howpublished}{\url{https://www.mitre.org/}}.
\newblock
\newblock
\shownote{Accessed: Jan-2022}.


\bibitem[Fowler(1999)]%
        {Fowler1999}
\bibfield{author}{\bibinfo{person}{Martin Fowler}.}
  \bibinfo{year}{1999}\natexlab{}.
\newblock \bibinfo{booktitle}{\emph{Refactoring: Improving the Design of
  Existing Code}}.
\newblock \bibinfo{publisher}{Addison-Wesley Longman Publishing Co., Inc.},
  \bibinfo{address}{Boston, MA, USA}.
\newblock
\showISBNx{0-201-48567-2}


\bibitem[Fucci et~al\mbox{.}(2021)]%
        {Fucci2021MSR}
\bibfield{author}{\bibinfo{person}{Gianmarco Fucci}, \bibinfo{person}{Nathan
  Cassee}, \bibinfo{person}{Fiorella Zampetti}, \bibinfo{person}{Nicole
  Novielli}, \bibinfo{person}{Alexander Serebrenik}, {and}
  \bibinfo{person}{Massimiliano Di~Penta}.} \bibinfo{year}{2021}\natexlab{}.
\newblock \showarticletitle{Waiting around or job half-done? Sentiment in
  self-admitted technical debt}. In \bibinfo{booktitle}{\emph{2021 IEEE/ACM
  18th International Conference on Mining Software Repositories (MSR)}}.
  \bibinfo{publisher}{IEEE/ACM}, \bibinfo{pages}{403--414}.
\newblock
\urldef\tempurl%
\url{https://doi.org/10.1109/MSR52588.2021.00052}
\showDOI{\tempurl}


\bibitem[GitHub(2022)]%
        {githubActions}
\bibfield{author}{\bibinfo{person}{Inc. GitHub}.}
  \bibinfo{year}{2022}\natexlab{}.
\newblock \bibinfo{title}{GitHub Actions}.
\newblock \bibinfo{howpublished}{\url{https://github.com/features/actions}}.
\newblock
\newblock
\shownote{Accessed: Jan-2022}.


\bibitem[Gkortzis et~al\mbox{.}(2018)]%
        {GkortzisEtAl2018}
\bibfield{author}{\bibinfo{person}{Antonios Gkortzis},
  \bibinfo{person}{Dimitris Mitropoulos}, {and} \bibinfo{person}{Diomidis
  Spinellis}.} \bibinfo{year}{2018}\natexlab{}.
\newblock \showarticletitle{VulinOSS: A Dataset of Security Vulnerabilities in
  Open-Source Systems}. In \bibinfo{booktitle}{\emph{Proceedings of the 15th
  International Conference on Mining Software Repositories, {MSR} 2018,
  Gothenburg, Sweden, May 28-29, 2018}}. \bibinfo{publisher}{ACM},
  \bibinfo{pages}{18--21}.
\newblock
\urldef\tempurl%
\url{https://doi.org/10.1145/3196398.3196454}
\showDOI{\tempurl}


\bibitem[Iammarino et~al\mbox{.}(2019)]%
        {Iammarino2019ICSME}
\bibfield{author}{\bibinfo{person}{Martina Iammarino},
  \bibinfo{person}{Fiorella Zampetti}, \bibinfo{person}{Lerina Aversano}, {and}
  \bibinfo{person}{Massimiliano Di~Penta}.} \bibinfo{year}{2019}\natexlab{}.
\newblock \showarticletitle{Self-Admitted Technical Debt Removal and
  Refactoring Actions: Co-Occurrence or More?}. In
  \bibinfo{booktitle}{\emph{2019 IEEE International Conference on Software
  Maintenance and Evolution (ICSME)}}. \bibinfo{publisher}{IEEE},
  \bibinfo{pages}{186--190}.
\newblock
\urldef\tempurl%
\url{https://doi.org/10.1109/ICSME.2019.00029}
\showDOI{\tempurl}


\bibitem[Li et~al\mbox{.}(2021)]%
        {LiFSE2021}
\bibfield{author}{\bibinfo{person}{Yi Li}, \bibinfo{person}{Shaohua Wang},
  {and} \bibinfo{person}{Tien~N. Nguyen}.} \bibinfo{year}{2021}\natexlab{}.
\newblock \bibinfo{booktitle}{\emph{Vulnerability Detection with Fine-Grained
  Interpretations}}.
\newblock \bibinfo{publisher}{Association for Computing Machinery},
  \bibinfo{address}{New York, NY, USA}, \bibinfo{pages}{292–303}.
\newblock
\showISBNx{9781450385626}
\urldef\tempurl%
\url{https://doi.org/10.1145/3468264.3468597}
\showURL{%
\tempurl}


\bibitem[Lim et~al\mbox{.}(2012)]%
        {LimEtAl2013}
\bibfield{author}{\bibinfo{person}{E. Lim}, \bibinfo{person}{N. Taksande},
  {and} \bibinfo{person}{C. Seaman}.} \bibinfo{year}{2012}\natexlab{}.
\newblock \showarticletitle{A Balancing Act: What Software Practitioners Have
  to Say about Technical Debt}.
\newblock \bibinfo{journal}{\emph{IEEE Software}} \bibinfo{volume}{29},
  \bibinfo{number}{6} (\bibinfo{date}{Nov} \bibinfo{year}{2012}),
  \bibinfo{pages}{22--27}.
\newblock
\showISSN{0740-7459}
\urldef\tempurl%
\url{https://doi.org/10.1109/MS.2012.130}
\showDOI{\tempurl}


\bibitem[Liu et~al\mbox{.}(2018)]%
        {LiuEtAl2018}
\bibfield{author}{\bibinfo{person}{Zhongxin Liu}, \bibinfo{person}{Qiao Huang},
  \bibinfo{person}{Xin Xia}, \bibinfo{person}{Emad Shihab},
  \bibinfo{person}{David Lo}, {and} \bibinfo{person}{Shanping Li}.}
  \bibinfo{year}{2018}\natexlab{}.
\newblock \showarticletitle{SATD Detector: A Text-Mining-Based Self-Admitted
  Technical Debt Detection Tool}. In \bibinfo{booktitle}{\emph{Proceedings of
  the 40th International Conference on Software Engineering: Companion
  Proceeedings}}. \bibinfo{publisher}{ACM}, \bibinfo{pages}{9–12}.
\newblock
\urldef\tempurl%
\url{https://doi.org/10.1145/3183440.3183478}
\showDOI{\tempurl}


\bibitem[Maldonado et~al\mbox{.}(2017)]%
        {MaldonadoEtAl2017a}
\bibfield{author}{\bibinfo{person}{E.~D.~S. Maldonado}, \bibinfo{person}{R.
  Abdalkareem}, \bibinfo{person}{E. Shihab}, {and} \bibinfo{person}{A.
  Serebrenik}.} \bibinfo{year}{2017}\natexlab{}.
\newblock \showarticletitle{An Empirical Study on the Removal of Self-Admitted
  Technical Debt}. In \bibinfo{booktitle}{\emph{2017 IEEE International
  Conference on Software Maintenance and Evolution (ICSME)}}.
  \bibinfo{publisher}{IEEE}, \bibinfo{pages}{238--248}.
\newblock
\urldef\tempurl%
\url{https://doi.org/10.1109/ICSME.2017.8}
\showDOI{\tempurl}


\bibitem[Mazuera-Rozo et~al\mbox{.}(2021)]%
        {mazuerarozoEtAl2021}
\bibfield{author}{\bibinfo{person}{Alejandro Mazuera-Rozo},
  \bibinfo{person}{Anamaria Mojica-Hanke}, \bibinfo{person}{Mario
  Linares-Vásquez}, {and} \bibinfo{person}{Gabriele Bavota}.}
  \bibinfo{year}{2021}\natexlab{}.
\newblock \showarticletitle{Shallow or Deep? An Empirical Study on Detecting
  Vulnerabilities using Deep Learning}. In
  \bibinfo{booktitle}{\emph{Proceedings of the 2021 IEEE/ACM 29th International
  Conference on Program Comprehension (ICPC)}}. \bibinfo{publisher}{IEEE/ACM},
  \bibinfo{pages}{276--287}.
\newblock
\urldef\tempurl%
\url{https://doi.org/10.1109/ICPC52881.2021.00034}
\showDOI{\tempurl}
\showeprint{2103.11940}~[cs.SE]


\bibitem[Michael(2011)]%
        {Howard}
\bibfield{author}{\bibinfo{person}{Howard Michael}.}
  \bibinfo{year}{2011}\natexlab{}.
\newblock \bibinfo{title}{Security Development Lifecycle (SDL) Banned Function
  Calls}.
\newblock
\newblock
\urldef\tempurl%
\url{http://msdn.microsoft.com/en-us/library/bb288454.aspx}
\showURL{%
\tempurl}


\bibitem[Mozilla(2022)]%
        {bugzilla}
\bibfield{author}{\bibinfo{person}{Mozilla}.} \bibinfo{year}{2022}\natexlab{}.
\newblock \bibinfo{title}{Bugzilla}.
\newblock
  \bibinfo{howpublished}{\url{https://bugzilla.mozilla.org/show_bug.cgi?id=1106067}}.
\newblock
\newblock
\shownote{Accessed: Jan-2022}.


\bibitem[Neuhaus et~al\mbox{.}(2007)]%
        {NeuhausEtAl2007}
\bibfield{author}{\bibinfo{person}{Stephan Neuhaus}, \bibinfo{person}{Thomas
  Zimmermann}, \bibinfo{person}{Christian Holler}, {and}
  \bibinfo{person}{Andreas Zeller}.} \bibinfo{year}{2007}\natexlab{}.
\newblock \showarticletitle{Predicting Vulnerable Software Components}. In
  \bibinfo{booktitle}{\emph{Proceedings of the 14th ACM Conference on Computer
  and Communications Security}} (Alexandria, Virginia, USA)
  \emph{(\bibinfo{series}{CCS '07})}. \bibinfo{publisher}{Association for
  Computing Machinery}, \bibinfo{address}{New York, NY, USA},
  \bibinfo{pages}{529–540}.
\newblock
\showISBNx{9781595937032}
\urldef\tempurl%
\url{https://doi.org/10.1145/1315245.1315311}
\showDOI{\tempurl}


\bibitem[Nord et~al\mbox{.}(2016)]%
        {NordEtAl2016}
\bibfield{author}{\bibinfo{person}{Robert~L. Nord}, \bibinfo{person}{Ipek
  Ozkaya}, \bibinfo{person}{Edward~J. Schwartz}, \bibinfo{person}{Forrest
  Shull}, {and} \bibinfo{person}{Rick Kazman}.}
  \bibinfo{year}{2016}\natexlab{}.
\newblock \showarticletitle{Can Knowledge of Technical Debt Help Identify
  Software Vulnerabilities?}. In \bibinfo{booktitle}{\emph{Proceedings of the
  9th USENIX Conference on Cyber Security Experimentation and Test}} (Austin,
  TX) \emph{(\bibinfo{series}{CSET'16})}. \bibinfo{publisher}{USENIX
  Association}, \bibinfo{address}{USA}, \bibinfo{pages}{1}.
\newblock


\bibitem[of~Standards and (NIST)(2022a)]%
        {julietEtAl}
\bibfield{author}{\bibinfo{person}{National~Institute of Standards} {and}
  \bibinfo{person}{Technology (NIST)}.} \bibinfo{year}{2022}\natexlab{a}.
\newblock \bibinfo{title}{Juliet test suite v1.3}.
\newblock
  \bibinfo{howpublished}{\url{https://samate.nist.gov/SRD/testsuite.php}}.
\newblock
\newblock
\shownote{Accessed: Jan-2022}.


\bibitem[of~Standards and (NIST)(2022b)]%
        {nvd}
\bibfield{author}{\bibinfo{person}{National~Institute of Standards} {and}
  \bibinfo{person}{Technology (NIST)}.} \bibinfo{year}{2022}\natexlab{b}.
\newblock \bibinfo{title}{National Vulnerabilities Database}.
\newblock \bibinfo{howpublished}{\url{https://nvd.nist.gov/}}.
\newblock
\newblock
\shownote{Accessed: Jan-2022}.


\bibitem[Potdar and Shihab(2014a)]%
        {PotdarShihab2014}
\bibfield{author}{\bibinfo{person}{Aniket Potdar} {and} \bibinfo{person}{Emad
  Shihab}.} \bibinfo{year}{2014}\natexlab{a}.
\newblock \showarticletitle{An Exploratory Study on Self-Admitted Technical
  Debt}. In \bibinfo{booktitle}{\emph{Proceedings of the 2014 IEEE
  International Conference on Software Maintenance and Evolution}}
  \emph{(\bibinfo{series}{ICSME '14})}. \bibinfo{publisher}{IEEE Computer
  Society}, \bibinfo{address}{USA}, \bibinfo{pages}{91–100}.
\newblock
\showISBNx{9781479961467}
\urldef\tempurl%
\url{https://doi.org/10.1109/ICSME.2014.31}
\showDOI{\tempurl}


\bibitem[Potdar and Shihab(2014b)]%
        {SATDpatterns}
\bibfield{author}{\bibinfo{person}{Aniket Potdar} {and} \bibinfo{person}{Emad
  Shihab}.} \bibinfo{year}{2014}\natexlab{b}.
\newblock \bibinfo{title}{List of SATD patterns}.
\newblock
  \bibinfo{howpublished}{\url{http://users.encs.concordia.ca/~eshihab/data/ICSME2014/satd.html}}.
\newblock
\newblock
\shownote{Accessed: Jan-2022}.


\bibitem[Ren et~al\mbox{.}(2019)]%
        {Ren2019TOSEM}
\bibfield{author}{\bibinfo{person}{Xiaoxue Ren}, \bibinfo{person}{Zhenchang
  Xing}, \bibinfo{person}{Xin Xia}, \bibinfo{person}{David Lo},
  \bibinfo{person}{Xinyu Wang}, {and} \bibinfo{person}{John Grundy}.}
  \bibinfo{year}{2019}\natexlab{}.
\newblock \showarticletitle{Neural Network-Based Detection of Self-Admitted
  Technical Debt: From Performance to Explainability}.
\newblock \bibinfo{journal}{\emph{ACM Trans. Softw. Eng. Methodol.}}
  \bibinfo{volume}{28}, \bibinfo{number}{3}, Article \bibinfo{articleno}{15}
  (\bibinfo{date}{jul} \bibinfo{year}{2019}), \bibinfo{numpages}{45}~pages.
\newblock
\showISSN{1049-331X}
\urldef\tempurl%
\url{https://doi.org/10.1145/3324916}
\showDOI{\tempurl}


\bibitem[repository(2022a)]%
        {cve}
\bibfield{author}{\bibinfo{person}{CVE repository}.}
  \bibinfo{year}{2022}\natexlab{a}.
\newblock \bibinfo{title}{Common Vulnerabilities and Enumerations ({CVE})}.
\newblock \bibinfo{howpublished}{\url{https://cve.mitre.org/index.html}}.
\newblock
\newblock
\shownote{Accessed: Jan-2022}.


\bibitem[repository(2022b)]%
        {cwe}
\bibfield{author}{\bibinfo{person}{CWE repository}.}
  \bibinfo{year}{2022}\natexlab{b}.
\newblock \bibinfo{title}{Common Weakness Enumeration ({CWE})}.
\newblock \bibinfo{howpublished}{\url{https://cwe.mitre.org/}}.
\newblock
\newblock
\shownote{Accessed: Jan-2022}.


\bibitem[repository(2022c)]%
        {chromiumRepo}
\bibfield{author}{\bibinfo{person}{Chromium repository}.}
  \bibinfo{year}{2022}\natexlab{c}.
\newblock \bibinfo{title}{Federally Funded Research and Development Centers}.
\newblock \bibinfo{howpublished}{\url{https://github.com/chromium/chromium}}.
\newblock
\newblock
\shownote{Accessed: Jan-2022}.


\bibitem[Russell et~al\mbox{.}(2018a)]%
        {russellEtAl2018}
\bibfield{author}{\bibinfo{person}{Rebecca~L. Russell}, \bibinfo{person}{Louis
  Kim}, \bibinfo{person}{Lei~H. Hamilton}, \bibinfo{person}{Tomo Lazovich},
  \bibinfo{person}{Jacob~A. Harer}, \bibinfo{person}{Onur Ozdemir},
  \bibinfo{person}{Paul~M. Ellingwood}, {and} \bibinfo{person}{Marc~W.
  McConley}.} \bibinfo{year}{2018}\natexlab{a}.
\newblock \bibinfo{title}{Automated Vulnerability Detection in Source Code
  Using Deep Representation Learning}.
\newblock
\newblock
\showeprint[arxiv]{1807.04320}~[cs.LG]


\bibitem[Russell et~al\mbox{.}(2018b)]%
        {VDISC}
\bibfield{author}{\bibinfo{person}{Rebecca~L. Russell}, \bibinfo{person}{Louis
  Kim}, \bibinfo{person}{Lei~H. Hamilton}, \bibinfo{person}{Tomo Lazovich},
  \bibinfo{person}{Jacob~A. Harer}, \bibinfo{person}{Onur Ozdemir},
  \bibinfo{person}{Paul~M. Ellingwood}, {and} \bibinfo{person}{Marc~W.
  McConley}.} \bibinfo{year}{2018}\natexlab{b}.
\newblock \bibinfo{title}{DRAPER VDISC dataset}.
\newblock \bibinfo{howpublished}{\url{https://osf.io/d45bw/wiki/home/}}.
\newblock
\newblock
\shownote{Accessed: Jan-2022}.


\bibitem[Russo et~al\mbox{.}(2022)]%
        {zenodo}
\bibfield{author}{\bibinfo{person}{Barbara Russo}, \bibinfo{person}{Matteo
  Camilli}, {and} \bibinfo{person}{Moritz Mock}.}
  \bibinfo{year}{2022}\natexlab{}.
\newblock \bibinfo{title}{Replication package}.
\newblock \bibinfo{howpublished}{\url{https://doi.org/10.5281/zenodo.5569313}}.
\newblock
\newblock
\shownote{Accessed: Jan-2022}.


\bibitem[Sabetta and Bezzi(2018)]%
        {SabettaBezzi2018}
\bibfield{author}{\bibinfo{person}{A. Sabetta} {and} \bibinfo{person}{M.
  Bezzi}.} \bibinfo{year}{2018}\natexlab{}.
\newblock \showarticletitle{A Practical Approach to the Automatic
  Classification of Security-Relevant Commits}. In
  \bibinfo{booktitle}{\emph{2018 IEEE International Conference on Software
  Maintenance and Evolution (ICSME)}}. \bibinfo{publisher}{IEEE Computer
  Society}, \bibinfo{address}{Los Alamitos, CA, USA},
  \bibinfo{pages}{579--582}.
\newblock
\urldef\tempurl%
\url{https://doi.org/10.1109/ICSME.2018.00058}
\showDOI{\tempurl}


\bibitem[security(2022)]%
        {exploitDB}
\bibfield{author}{\bibinfo{person}{Offensive security}.}
  \bibinfo{year}{2022}\natexlab{}.
\newblock \bibinfo{title}{Exploit database}.
\newblock \bibinfo{howpublished}{\url{https://www.exploit-db.com/}}.
\newblock
\newblock
\shownote{Accessed: Jan-2022}.


\bibitem[Siavvas et~al\mbox{.}(2019)]%
        {siavvas2019empirical}
\bibfield{author}{\bibinfo{person}{Miltiadis Siavvas},
  \bibinfo{person}{Dimitrios Tsoukalas}, \bibinfo{person}{Marija Jankovic},
  \bibinfo{person}{Dionysios Kehagias}, \bibinfo{person}{Alexander
  Chatzigeorgiou}, \bibinfo{person}{Dimitrios Tzovaras}, \bibinfo{person}{Nenad
  Anicic}, {and} \bibinfo{person}{Erol Gelenbe}.}
  \bibinfo{year}{2019}\natexlab{}.
\newblock \showarticletitle{An empirical evaluation of the relationship between
  technical debt and software security}. In \bibinfo{booktitle}{\emph{9th
  International Conference on Information society and technology (ICIST)}},
  Vol.~\bibinfo{volume}{2019}.
\newblock


\bibitem[Sierra et~al\mbox{.}(2019)]%
        {SierraEtAl2019}
\bibfield{author}{\bibinfo{person}{Giancarlo Sierra}, \bibinfo{person}{Emad
  Shihab}, {and} \bibinfo{person}{Yasutaka Kamei}.}
  \bibinfo{year}{2019}\natexlab{}.
\newblock \showarticletitle{A survey of self-admitted technical debt}.
\newblock \bibinfo{journal}{\emph{Journal of Systems and Software}}
  \bibinfo{volume}{152} (\bibinfo{year}{2019}), \bibinfo{pages}{70--82}.
\newblock
\showISSN{0164-1212}
\urldef\tempurl%
\url{https://doi.org/10.1016/j.jss.2019.02.056}
\showDOI{\tempurl}


\bibitem[Stuckman et~al\mbox{.}(2017)]%
        {StuckmanEtAl2017}
\bibfield{author}{\bibinfo{person}{Jeffrey Stuckman}, \bibinfo{person}{James
  Walden}, {and} \bibinfo{person}{Riccardo Scandariato}.}
  \bibinfo{year}{2017}\natexlab{}.
\newblock \showarticletitle{The Effect of Dimensionality Reduction on Software
  Vulnerability Prediction Models}.
\newblock \bibinfo{journal}{\emph{{IEEE} Trans. Reliab.}} \bibinfo{volume}{66},
  \bibinfo{number}{1} (\bibinfo{year}{2017}), \bibinfo{pages}{17--37}.
\newblock
\urldef\tempurl%
\url{https://doi.org/10.1109/TR.2016.2630503}
\showDOI{\tempurl}


\bibitem[Ward(2009)]%
        {Cunningham2009}
\bibfield{author}{\bibinfo{person}{Cunningham Ward}.}
  \bibinfo{year}{2009}\natexlab{}.
\newblock \bibinfo{title}{Ward explains Debt Metaphor}.
\newblock
\newblock
\urldef\tempurl%
\url{wiki.c2.com/?WardExplainsDebtMetaphor}
\showURL{%
\tempurl}


\bibitem[Wehaibi et~al\mbox{.}(2016)]%
        {WehaibiEtAl2016}
\bibfield{author}{\bibinfo{person}{S. Wehaibi}, \bibinfo{person}{E. Shihab},
  {and} \bibinfo{person}{L. Guerrouj}.} \bibinfo{year}{2016}\natexlab{}.
\newblock \showarticletitle{Examining the Impact of Self-Admitted Technical
  Debt on Software Quality}. In \bibinfo{booktitle}{\emph{2016 IEEE 23rd
  International Conference on Software Analysis, Evolution, and Reengineering
  (SANER)}}, Vol.~\bibinfo{volume}{1}. \bibinfo{publisher}{IEEE},
  \bibinfo{pages}{179--188}.
\newblock
\urldef\tempurl%
\url{https://doi.org/10.1109/SANER.2016.72}
\showDOI{\tempurl}


\bibitem[Yu et~al\mbox{.}(5555)]%
        {ZYu2020TSE}
\bibfield{author}{\bibinfo{person}{Z. Yu}, \bibinfo{person}{F. Fahid},
  \bibinfo{person}{H. Tu}, {and} \bibinfo{person}{T. Menzies}.}
  \bibinfo{year}{5555}\natexlab{}.
\newblock \showarticletitle{Identifying Self-Admitted Technical Debts with
  Jitterbug: A Two-Step Approach}.
\newblock \bibinfo{journal}{\emph{IEEE Transactions on Software Engineering}}
  \bibinfo{number}{01} (\bibinfo{date}{oct} \bibinfo{year}{5555}),
  \bibinfo{pages}{1--1}.
\newblock
\showISSN{1939-3520}
\urldef\tempurl%
\url{https://doi.org/10.1109/TSE.2020.3031401}
\showDOI{\tempurl}


\bibitem[Zampetti et~al\mbox{.}(2018)]%
        {ZampettiEtAl2018}
\bibfield{author}{\bibinfo{person}{Fiorella Zampetti},
  \bibinfo{person}{Alexander Serebrenik}, {and} \bibinfo{person}{Massimiliano
  Di~Penta}.} \bibinfo{year}{2018}\natexlab{}.
\newblock \showarticletitle{Was Self-admitted Technical Debt Removal a Real
  Removal?: An In-depth Perspective}. In \bibinfo{booktitle}{\emph{Proceedings
  of the 15th International Conference on Mining Software Repositories, {MSR}
  2018, Gothenburg, Sweden, May 28-29, 2018}} (Gothenburg, Sweden)
  \emph{(\bibinfo{series}{MSR '18})}. \bibinfo{publisher}{ACM},
  \bibinfo{address}{New York, NY, USA}, \bibinfo{pages}{526--536}.
\newblock
\showISBNx{978-1-4503-5716-6}
\urldef\tempurl%
\url{https://doi.org/10.1145/3196398.3196423}
\showDOI{\tempurl}


\bibitem[Zazworka et~al\mbox{.}(2011)]%
        {ZazworkaEtAl2011}
\bibfield{author}{\bibinfo{person}{Nico Zazworka}, \bibinfo{person}{Michele~A.
  Shaw}, \bibinfo{person}{Forrest Shull}, {and} \bibinfo{person}{Carolyn
  Seaman}.} \bibinfo{year}{2011}\natexlab{}.
\newblock \showarticletitle{Investigating the Impact of Design Debt on Software
  Quality}. In \bibinfo{booktitle}{\emph{Proceedings of the 2nd Workshop on
  Managing Technical Debt}} (Waikiki, Honolulu, HI, USA)
  \emph{(\bibinfo{series}{MTD '11})}. \bibinfo{publisher}{Association for
  Computing Machinery}, \bibinfo{address}{New York, NY, USA},
  \bibinfo{pages}{17–23}.
\newblock
\showISBNx{9781450305860}
\urldef\tempurl%
\url{https://doi.org/10.1145/1985362.1985366}
\showDOI{\tempurl}


\bibitem[Zou et~al\mbox{.}(2019)]%
        {ZouEtAl2019}
\bibfield{author}{\bibinfo{person}{Deqing Zou}, \bibinfo{person}{Sujuan Wang},
  \bibinfo{person}{Shouhuai Xu}, \bibinfo{person}{Zhen Li}, {and}
  \bibinfo{person}{Hai Jin}.} \bibinfo{year}{2019}\natexlab{}.
\newblock \showarticletitle{$\mu$VulDeePecker: A Deep Learning-Based System for
  Multiclass Vulnerability Detection}.
\newblock \bibinfo{journal}{\emph{IEEE Transactions on Dependable and Secure
  Computing}}  \bibinfo{volume}{18} (\bibinfo{year}{2019}),
  \bibinfo{pages}{2224--2236}.
\newblock
\showISSN{2160-9209}
\urldef\tempurl%
\url{https://doi.org/10.1109/tdsc.2019.2942930}
\showDOI{\tempurl}


\end{thebibliography}

\end{document}